\documentclass[pre,showpacs,preprintmumbers,showkeys]{revtex4-1}
\usepackage{graphicx,amssymb,color}
\usepackage[dvipsnames]{xcolor}
\usepackage[normalem]{ulem}
\usepackage{color}
\begin{document}

\title{Surface waves propagation on a turbulent flow forced electromagnetically }
\author{Pablo Guti\'errez$^{1,2}$ \& S\'ebastien Auma\^\i tre$^{1,3}$}
\email[Corresponding author. Email address: ]{sebastien.aumaitre@cea.fr}
\affiliation{$^1$Service de Physique de l'Etat Condens\'e, DSM, CEA-Saclay, CNRS URA 2464, 91191 Gif-sur-Yvette, France\\
$^2$Departamento de F\'\i sica, Facultad de Ciencias F\' \i sicas y Matem\'aticas, Universidad del Chile, Avenida Blanco Encalada 2008, Santiago, Chile\\
$^3$Laboratoire de Physique, ENS de Lyon, UMR-CNRS 5672, 46 all\'ee d'Italie, F69007 Lyon, France  }

\date{\today}
\begin{abstract}
We study the propagation of  monochromatic surface waves on a turbulent flow. The flow is generated in a layer of liquid metal by an electromagnetic forcing. This forcing creates a quasi two-dimensional (2D) turbulence with strong vertical vorticity. The turbulent flow contains much more energy than the surface waves. In order to focus on the surface wave, the deformations induced by the turbulent flow are removed. This is done by performing a coherent phase averaging. For wavelengths smaller than the forcing lengthscale, we observe a significant increase of the wavelength of the propagating wave that has not been reported before. We suggest that it can be explained by the random deflection of the wave induced by the velocity gradient of the turbulent flow. Under this assumption, the wavelength shift is an estimate of the  fluctuations of deflection angle. The local measurements of the wave frequency far from the wavemaker do not reveal such systematic behavior, although a small shift is measured. Finally we quantify the damping enhancement  induced by the turbulent flow. We review various theoretical scaling laws proposed previously. Most of them propose a damping that increases as the square of  Froude number. In contrast, our experimental results show a turbulent damping increasing linearly with the Froude number. We interpret this linear behaviour as a balance between the time spent by a wave to cross a turbulent structure with the turbulent mixing time. The larger is the ratio of these 2 times, the more energy is extracted from the progressive wave. Finally, mechanisms of energy exchange and open issues are discussed and further studies  are proposed.
\end{abstract}

\maketitle

\section{Introduction}

Owing to their interest to probe the properties of complex matter, the various regimes of waves propagation through random media are the subject of intense studies \cite{Sheng,Ishimara}. Here, we consider surface waves, which are both dispersive (see Equation \ref{reladisp}) and nonlinear, propagating on a turbulent flow. They can be scattered by the velocity gradients and can exchange energy with the underlying flow. This is an issue for the energy exchanges between atmosphere and ocean and for the prediction of coastal swell. Because of the complexity of these phenomena, the interactions between wind-generated gravity waves and the underlying turbulent flow are not fully understood. Hence, they motivate many theoretical studies and in situ measurements in physical oceanography (see some examples in \cite{Phillips77,Kitaigorodskii1,Ardhuin2006,Veron2009} ).

The scattering of monochromatic waves by a single vortex has been studied both experimentally and theoretically \cite{VivancoPRE04,CostePRE99}.  Some similarities with the Aharonov-Bohm effect have been underlined. Some scaling laws have been proposed for the damping of waves by turbulence \cite{Lighthill53,PhillipsJFM59,Boyev71,TeixeiraJFM02,Kantha2006}. Lighthill \cite{Lighthill53} studied theoretically the elastic scattering of a sound wave by a turbulent flow. Phillips \cite{PhillipsJFM59} considered the same problem for gravity waves. He developed a scaling argument for wavelengths, $\lambda$, smaller than the characteristic length, $L$, of the turbulent forcing. Fabrikant and Raevsky computed the cross section of a  single vortex and applied it to gravity waves propagating on  a  turbulent drift flow \cite{FabrikantRaevskyJFM93}. Teixeira and Belcher \cite{TeixeiraJFM02} and Kantha \cite{Kantha2006} considered energy exchange due to the stretching of the vortices by the waves. In that case they assumed that the gravity waves have wavelength larger than $L$. As it will be underlined in section \ref{ThPred}, the point of view of Phillips, Fabrikant and Raevsky on the one hand, and the one of Teixeira and Belcher on the other, obtain the same scaling, although both are built on different physical mechanisms.  Indeed both approaches involve an increase of turbulent damping proportional to the square of Froude number, $Fr$, defined as the ratio of the flow velocity over the wave speed. By considering a more energetic underlying flow, Boyev \cite{Boyev71} proposed another scaling law, linear in $Fr$. Numerical studies are not easy because they merge the difficulties to simulate turbulence with those to model numerically a  realistic free interface. Guo and Shen devoted a laudable effort to solve these issues \cite{GuoShen2010,GuoShen2014}, but up to now only a small range of parameters has been explored \cite{GuoShen2014}. 

There are previous experimental studies of surface waves on a turbulent flow. Falcon and Fauve \cite{Claudio} studied the threshold of the Faraday instability that increases in the presence of a chaotic underlying flow driven by an electromagnetic forcing. Modifications of the wave's statistical properties are also exhibited. Green {\it et al} \cite{GeenNatPhysSci72} considered a progressive wave on a flow generated by a vibrating grid. They suggest a turbulent dissipative decay proportional to the square of the wave amplitude. Using a similar device, \"Olmez and Milgram \cite{OlmezJFM92} proposed a temporal damping due to turbulence, directly proportional to the turbulent mixing rate. They used arguments similar to those presented in \cite{Boyev71}. However, this result was not confirmed by Ermakov {\it et al} \cite{Ermakov2014}. By using the threshold of the Faraday instability to determine the damping induced by a flow generated by a vibrating grid, these authors show a damping in agreement with \cite{TeixeiraJFM02}.
Note that the experiments generating turbulence with an oscillating grid introduce another frequency that can interfere with the one of the waves. Moreover, using mainly local probes, these experiments cannot study the wavelength. Finally, the action of a spatiotemporal noise on the Faraday instability is by itself a complex noise-induced phenomenon. 

Our experimental device combines the electromagnetic driving used in \cite{Claudio} to stir the flow and progressive waves generated by a paddle. By the use of a liquid alloy, Gallinstan \cite{Gallinstan}, a strong Electromagnetic Driven Flow (EMDF) can be supplied. The high electrical currents do not heat the fluid by Joule effect. This EMDF produces rapid random fluctuations in the bulk of the fluid. Moreover, the waves generated at  the surface are less damped in Gallinstan than in water, because of its lower dynamic viscosity  (2.5 times smaller than water). By means of this EMDF in a liquid metal, the turbulent energy is large compared to the energy carried by the waves. Finally, by using the diffusion of a laser sheet on the surface and triangulation techniques, we are able to follow the surface elevation along a line in the direction of the wave propagation. The random fluctuations of the surface are removed by a coherent average procedure. Hence we can measure the wavelength of the propagating wave with and without EMDF and the damping induced by turbulence. 

 In the following section \ref{ExpFeat}, we expose the main features of the experiment. The setup is described and some estimates of relevant dimensionless parameters are given. The main features of the EMDF and the generated waves are also described. Next, in the section \ref{Shift}, we focus on the increase of the wavelength induced by the growth of the turbulent flow when the wave frequency is large. It is interpreted in terms of a random fluctuation of the direction of the wave propagation. Section \ref{Damp} concerns the enhancement of the damping by the turbulence. In order to interpret the data, we resume some theoretical scalings, mainly based on dimensional arguments, and we confront them to our experimental findings. Our data suggest a linear scaling with $Fr$. This scaling may be interpreted by the horizontal mixing rate of the wave. The physical damping mechanism, as well as further works to perform, are discussed in the concluding section.


\section{Experimental Features}
\label{ExpFeat}

\subsection{The setup}

\subsubsection{Turbulent flows and wave generation} 

\begin{figure}
\centering
\begin{minipage}[c]{0.4\columnwidth} 
\includegraphics[width= 1\columnwidth,angle=0]{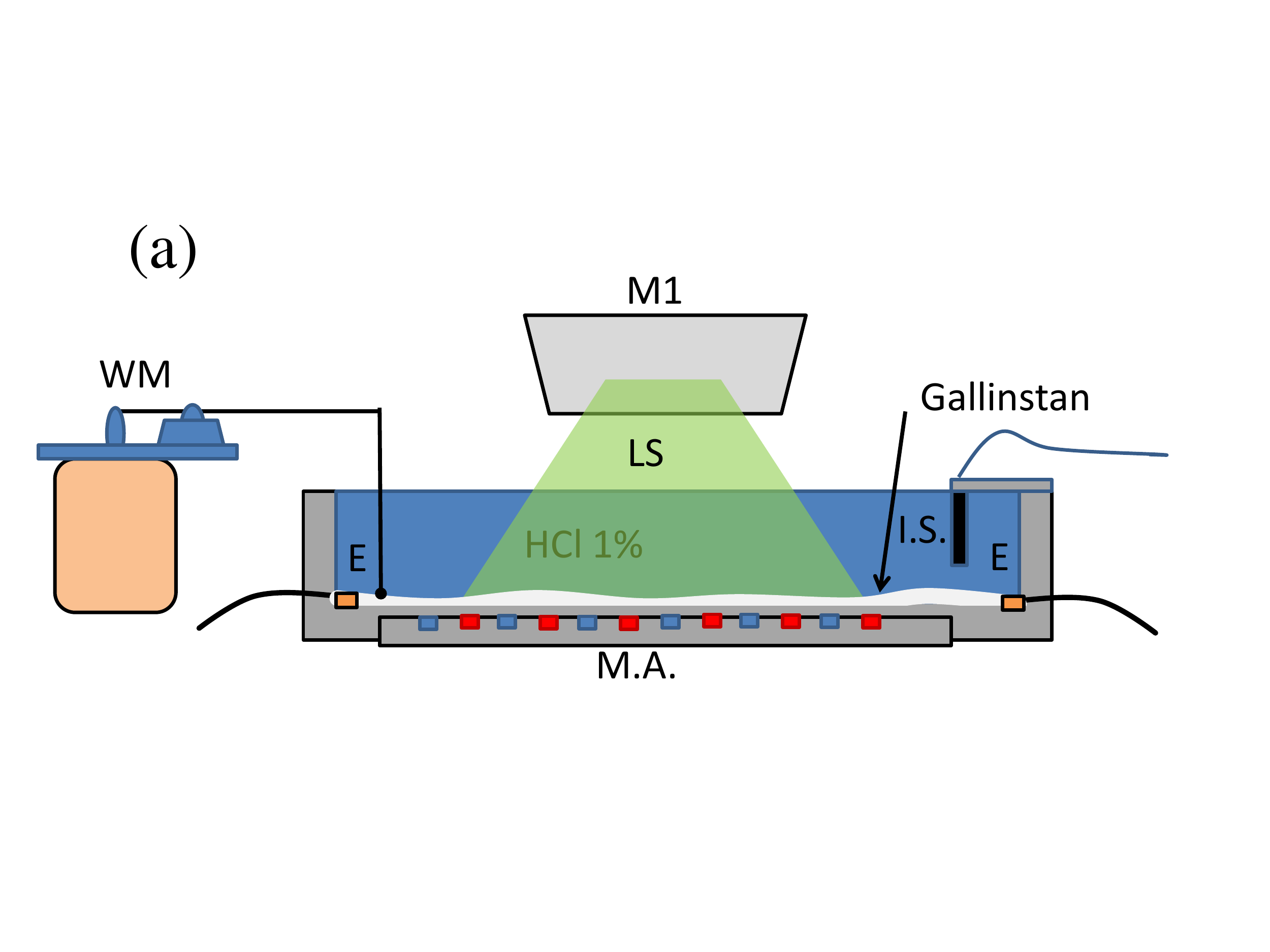}

\vspace{-3cm}
\includegraphics[width= .6\columnwidth,angle=0]{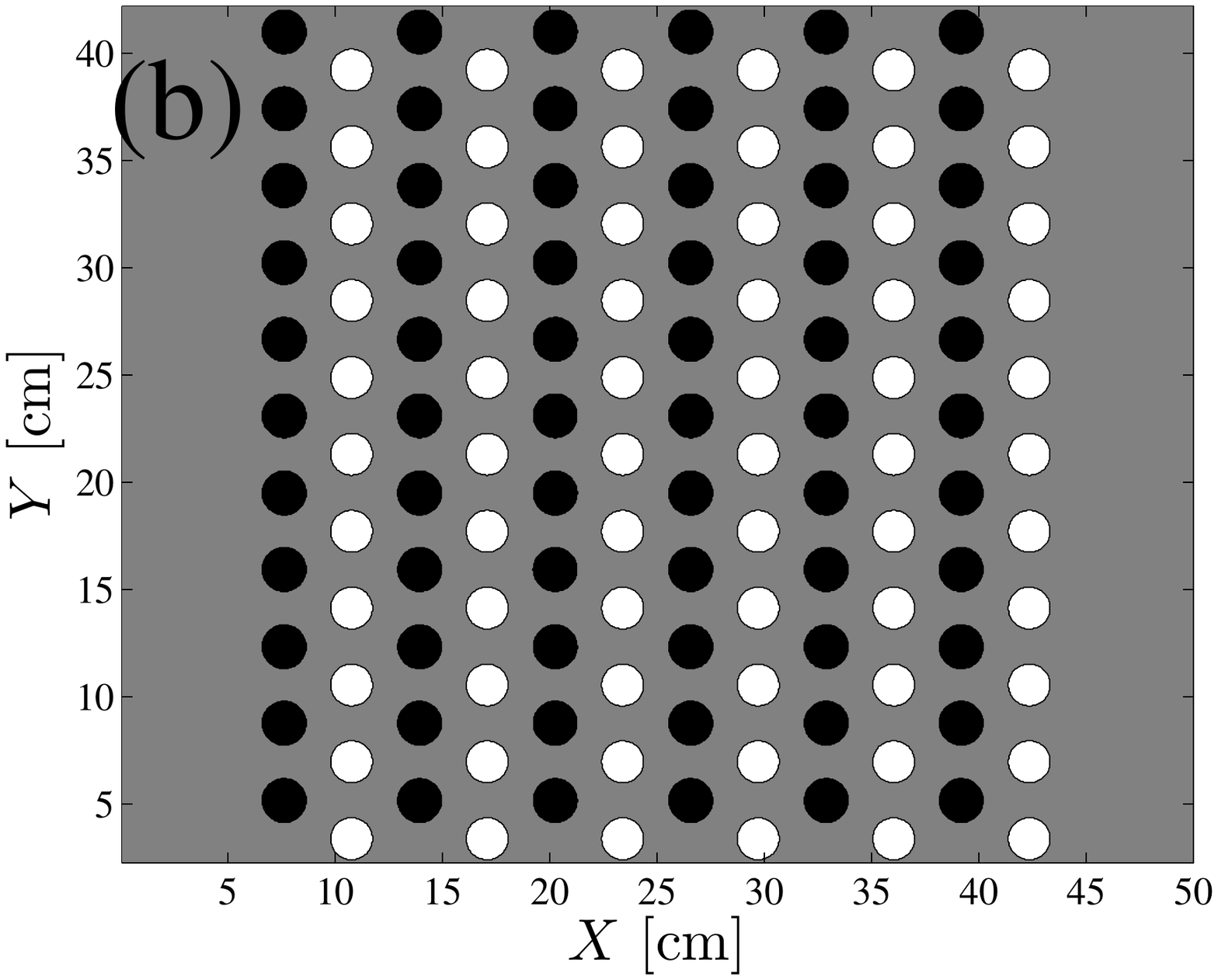}
\end{minipage} 
\begin{minipage}[c]{0.4\columnwidth} 
(c)
\includegraphics[width= 1\columnwidth,angle=0]{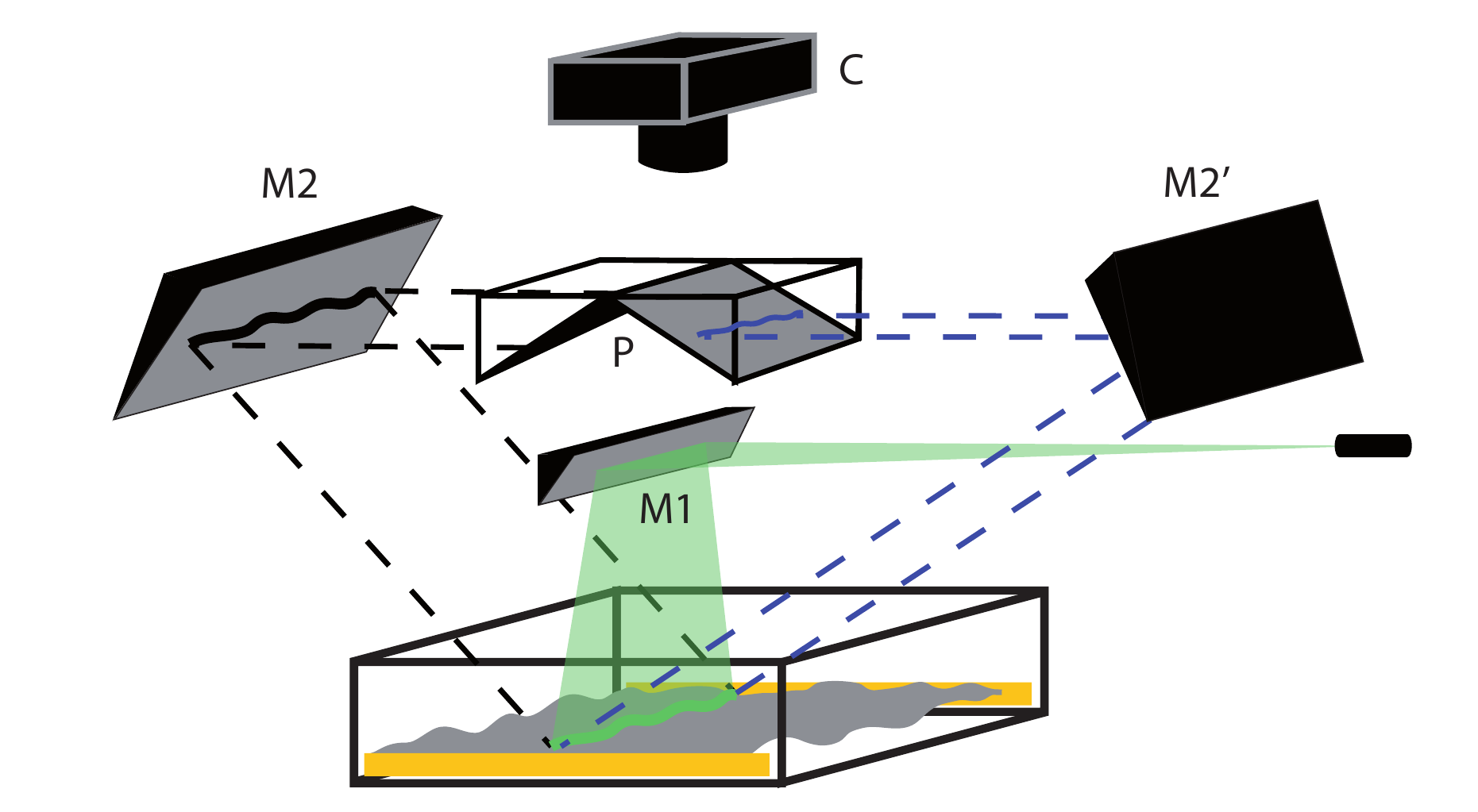}
\end{minipage}
\vspace{-1cm}
\caption{Experimental device. {\bf (a)}: a cut along the wave propagation axis with WM the Wavemaker, LS the Laser Sheet, MA the Magnet Array, IS the Inductive Sensor, and E the Electrodes. {\bf (b):} The structure of the magnets array beneath the cell. {\bf (c)}: geometry of the device to track the diffused line. The laser sheet is sent to the surface with the mirror M1. The diffused line is imaged twice on the camera C with two opposite angle by the mirror M2 and M2'. Each image is caught on half of the camera sensor by using 2 prisms P.} 
\label{CellArtII}
\end{figure}

The device driving the turbulent flow has been previously described in \cite{GutierrezTh}. It is depicted in figure \ref{CellArtII}. It consists of a rectangular tank of size $0.5\times 0.4\times0.1$ m$^3$. It is filled with a layer of depth  $H=1$  cm of Gallinstan covered by a layer of water slightly acidified to prevent the oxidation of Gallinstan surface. The water layer is thick  enough (about 8 cm) to make the water--Gallinstan interface insensitive to the boundary condition at top of the water layer (covered by a Plexiglas plate). Therefore, we study the deformation of a Gallinstan-water interface. Note that the Gallinstan is 6.5 times denser than water. We use the Lorentz force to stir the fluid. This force is generated by a horizontal current and  an array of permanent disc magnets of opposite polarities, depicted in figure\ref{CellArtII}--b. The distance between magnets defines the forcing length scale $l=3.7$ cm.  The magnetic field generated at a magnet center  reaches $B=0.12$ T at  the bottom of the liquid metal layer, but it is reduced to 0.033 T at the liquid metal surface. The supplied current $I$ ranges between 0 and 180 A, for most of the experiments presented here. $I$ is our control paramater driving the velocity gradients and the turbulent strength, via the Lorentz force. 

\begin{figure}
\centering
\vspace{-4cm}
\includegraphics[width= 0.9\columnwidth,angle=0]{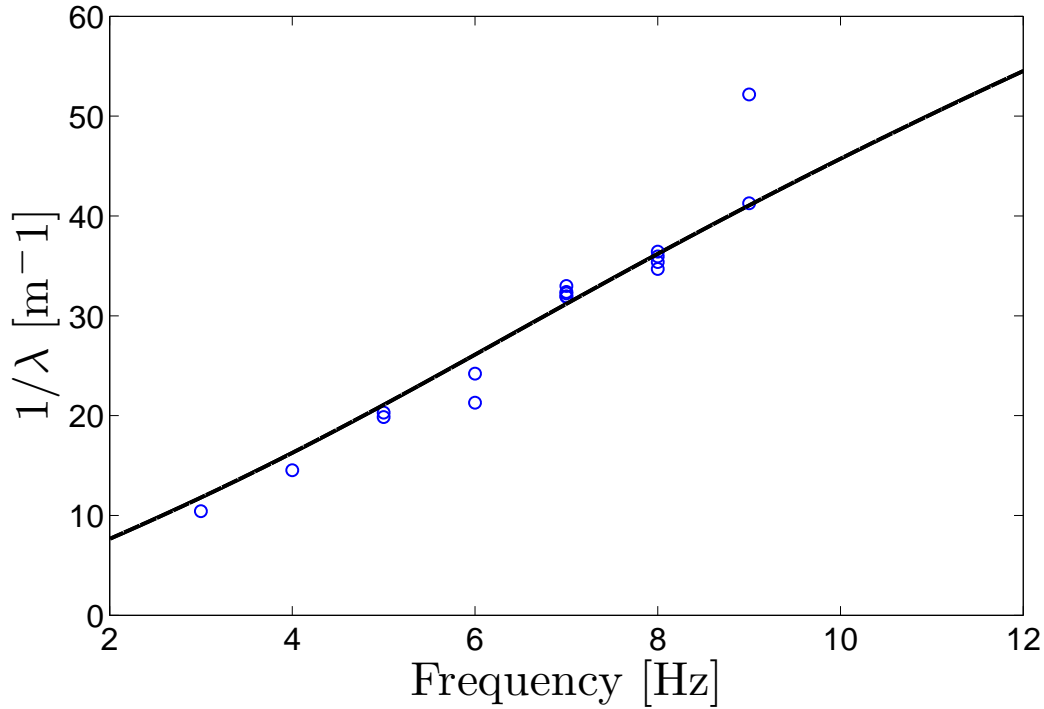}
\vspace{-4cm}
\caption{Dispersion relation without turbulent flow. Circles correspond to experimental measurements. The full line is the prediction of Equation (\ref{reladisp}) with $\sigma=0.5$ N/m. }
\label{DispRel}
\end{figure}

The waves are excited by a cylindrical paddle of diameter 8 mm around a horizontal axis and length 100 mm, in contact with the interface (see figure \ref{CellArtII}). An electromagnetic shaker LDS 406 moves it vertically. In order to prevent strong nonlinearties and wave breaking, the paddle displacement is limited at most to 10\% of the wavelength for the present set of experiments. Hence it never overcomes 6 mm for the longest generated  wavelength.  By pushing the fluid surface upward and downward it creates propagating waves without the generation of  stream. The direction of propagation defines the x-axis. It is parallel to the applied electrical current. We drive the waves at frequencies $\nu$ between 3 and 9 Hz. The dispersion relation for the waves between two fluids is given by :
\begin{equation}
\omega^2=\left(\frac{\rho-\rho'}{\rho+\rho'}g k+\frac{\sigma}{\rho+\rho'}k^3\right)\cdot\tanh( k H),
\label{reladisp}
\end{equation}
where $k=2\pi/\lambda$ is the wavenumber, $\omega=2\pi\nu$ is the corresponding angular frequency, $H$ is the fluid depth, $g$ is the gravity constant, $\sigma$ is the surface tension, $\rho$ is Gallinstan density and $\rho'$ is the acidified--water density. The phase velocity, $C_w =\omega/k$ and the group velocity $C_g=d\omega/dk$ differ in general. The limit of {\it deep water} is reached for $H/\lambda>1$, hence $\tanh( k H)\sim 1$. The right hand side of relation (\ref{reladisp}) reduces to a polynomial in this case. Capillarity  dominates for $k\gg1/l_c$ with $l_c=\sqrt{\sigma/(\rho g)}=2.8$ mm, the capillary length. Elsewhere, gravity is the main restoring force. In deep water, when only one of the restoring forces dominates, the dispertion relation  (\ref{reladisp}) simplifies to a power law. Unfortunately, Tab. 1 shows that we are in none of these limits. Therefore we have to use the full dispersion relation (\ref{reladisp}) to compute $C_w$ or $C_g$. Note that $C_g$ is nearly constant. It is because our range of forcing frequencies is around its minimum (reached at 6.37 Hz). The surface tension between Gallinstan and water is not reported and has to be estimated. Moreover it might depend on the acid concentration (about 1\% in our experiment). By using the Faraday instability, we determined a surface tension of  $0.5$ N/m under our experimental conditions. The dispersion relation represented by the solid line on figure \ref{DispRel} is obtained for this value of $\sigma$.  We use mainly this value hereafter. However the surface tension is very sensitive to the presence of impurities. Therefore, we adjust it  in some experiments to take into account the aging of the interface that explains the scattering of the experimental data on figure \ref{DispRel}. The correction never exceeded 20\% of the nominal value (0.5 N/m).

\begin{center}
\begin{tabular}{|c|c|c|c|c|c|c|c|}

\hline 
$\nu$ Hz &3 &4 &  5 &  6&  7&   8 &  9\\
\hline 
$\lambda$ cm &8.48& 6.15& 4.74&3.83&3.21&2.76&2.44\\
\hline 
tanh$(kH)$& 0.63&0.77&0.87&0.93& 0.96&0.98& 0.99\\
\hline
$C_w$ (m/s)&0.25&0.25 &0.24&0.23&0.22&0.22 &0.22\\
\hline
$C_g$(m/s)&0.23&0.21&0.20&0.20&0.20&0.20&0.21\\
\hline
$C_w/C_g$&1.09&1.14&1.17&1.17 &1.14&1.09&1.04\\
\hline
$Fr^{max}_w=U_o/C_w$&0.65& 0.68&0.70 &0.72 &0.74 &0.75&0.76\\
\hline
$Fr^{max}_g=U_o/C_g$&0.71&0.77&0.82&0.84&0.84&0.82&0.79\\
\hline
\end{tabular}
\medskip
\noindent
\end{center}
Tab. 1 Main features of the excited wave. The characteristic velocity of the flow $U_o$ is estimated for a typical applied current of 160 A.\\

\subsubsection{Measurement technics and averaging procedure }

We use inductive sensors to measure the wavemaker position and to probe locally the surface. The measurement of the surface along the direction of propagation is sketched on figure \ref{CellArtII}--c. A laser sheet is sent vertically on the surface. The deformation of the line diffused by the surface gives the deformation of the surface by a triangulation method. The diffusion of the laser light is quite small due to the high reflectivity of the Gallinstan surface. Therefore high sensitivity is needed. However the surface deformation can imply some direct reflections in the direction of the camera sensor. This can blind the diffused line. Therefore, two simultaneous pictures at different angles are necessary to follow the deformation of all the line at all time. Indeed the diffused line cannot be blinded in the same time at the same position on both images. Instead of using two cameras, we built the optical setup shown in figure \ref{CellArtII}--right. It allows us to gather two simultaneous pictures at  angles of $\pm 37 ^o$ with the vertical  with the use of a single sensitive camera. An algorithm tracks the diffused line, which is assumed continuous and derivable.

In order to remove all the turbulent deformations that are not in phase with the driving, we perform a coherent phase average as suggested in \cite{Kitaigorodskii2}. To do so, we take precisely $n$ images per wave period ($5\leq n \leq15$). We perform an average over more than 100 periods of all detected lines that are exactly locked in phase with the excited wave. Therefore we get $n$ spatially averaged profiles  for various turbulent strengths of the EMDF. We reach a precision of the 0.25 mm for the wavelength and of 0.15 mm for the wave amplitude.

\subsection{Dimensionless parameters}
\label{dimensionless}

The dimensionless Navier--Stokes Equation, driven by an electromagnetic Lorentz force, exhibits a natural velocity scale $U_o=\sqrt{JB l/\rho}$, which balances the advection term and the Lorentz force. Here, we use the forcing length $l$ as the characteristic length of the flow. Thus, one gets the Reynolds number $Re=\sqrt{JB/\rho}\cdot l^{3/2}/\nu_G$, with $\nu_G$ the dynamic viscosity of Gallinstan. Taking the maximum value of the magnetic field  $B=0.12$ T, we can expect $ U_o\sim 16 $ cm/s for $I=180$ A, and thus $Re=1.5\times10^4$, in our device. These estimates are compared to experimental measurements in the next section. Note that the Re here is 10 times larger than the one reached with grid generated flows
\cite{GeenNatPhysSci72,OlmezJFM92,Ermakov2014}. Such estimates of the Reynolds number give a Kolmogorov length, $ l_K=\rho \eta^{3/4}/\epsilon^{1/4}\sim 3\times 10^{-2}$ mm with $\epsilon\sim U_o^3/L$ the energy flux by unit of mass. In a thin fluid layer, the friction on the bottom induces a damping of the velocity. This friction term acts at all scales and modifies the scalings of the two-dimensional (2D) turbulence \cite{Young}. In a liquid metal submitted to an electromagnetic forcing, friction is concentrated to a thin magnetic boundary layer where the electromagnetic induction phenomena focus the electrical currents and the velocity gradients \cite{Sommeria_1986}. The depth of this layer $e_H=H/H\!a$, is characterized by the Hartmann number $H\!a=\sqrt{\sigma/\eta}BH$ \cite{Gallinstan}. Hence $e_H$ can be as small as 0.2 mm just above the magnets. One can evaluate this friction strength by the Reynolds number built on Hartmann layer $Re_H=Re_L/H\!a\cdot H/L=\sqrt{JlR/\sigma\nu B}\sim 100$. Therefore the system generates a highly anisotropic nonlinear flow with mainly vertical vorticity.

A major dimensionless parameter in the study of the interaction between wave and flow is the ratio of the flow velocity over the wave velocity. For acoustic non dispersive waves, it defines the Mach number. For dispersive surface waves, it is called Froude number. One can distinguish between a Froude number based on the phase velocity of the wave $Fr_w=U_o/C_w$ and another one based on the group velocity of the waves $Fr_g=U_o/C_g$ . Note that the Froude numbers introduced here are built with the complete gravito-capillary wave velocity. Hence, it includes capillary effects. As mentioned previously, Tab. 1 shows that none of these effects are negligible in our experiment. The Froude number based on the root mean square (RMS) velocity will be smaller about a factor 0.4 than those shown in Tab. 1. Nevertheless, we reach larger Froude numbers than in previous experiments \cite{Claudio,GeenNatPhysSci72,OlmezJFM92,Ermakov2014}. Actually the energy by unit volume contained in the turbulence, $E_t=\rho\langle U^2\rangle$, is much larger than the energy of the generated wave, $E_w=\rho h_o^2\omega^2$, with $h_o$ the wave amplitude. The ratio between these energies can be expressed as $E_t/E_w=(Fr_w\cdot\lambda/h _o)^2$. Here, we keep the wave steepness $h _o/\lambda$ less than 0.1 to prevent wave breaking. Therefore, the energy ratio is larger than 10 in most of our parameter range.  Hence, one may  expect stronger interaction and new interesting effects of the flow on the waves.

\subsection{General features of the EMDF }

\begin{figure}
\centering
\vspace{-4cm}
\includegraphics[width= 0.9\columnwidth,angle=0]{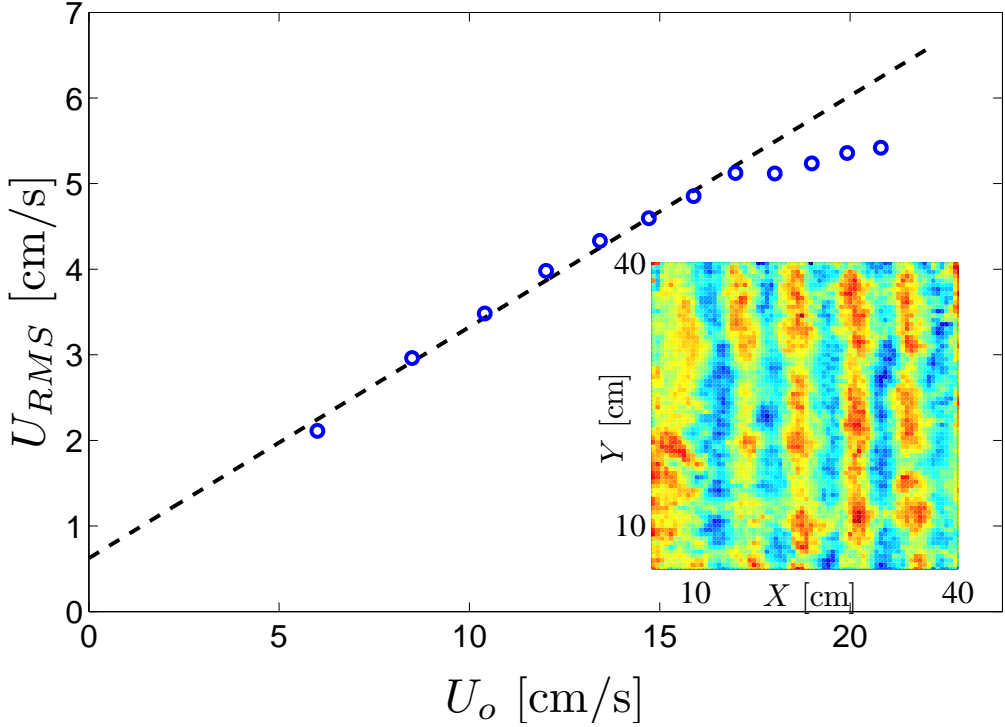}
\vspace{-4cm}
\caption{{\bf Main panel:}  RMS velocity of floaters tracked at the liquid metal surface, $U_{RMS}$, as a function of the charateristic MHD velocity $U_o=\sqrt{JB l/\rho}$. The linear fit (dashed line) has a slope of: 0.27. {\bf Inset:}  the spatial distribution of vertical vorticity averaged over 60 s, for a driving current of I =  150 A ($U_o$= 14.7 cm/s). The color map extends from -10 s$^{-1}$ (blue) to 10 s$^{-1}$ (red).}
\label{velocity}
\end{figure}

We give here some features of the EMDF, as detailed in [22]. In order to estimate the velocity field without waves, we seed the surface with millimetric particles of density $\rho_p = $ 2000 kg/m$^{3}$. The particles are lighter than Galinstan ($\rho_p/\rho \sim 0.3$) [21], and heavier than water ($\rho_p/\rho' \sim 2$), hence, they are floating in Galinstan, always at its interface with water. Using usual technics of particle tracking velocimetry, we measure the horizontal velocity of these floaters. In figure \ref{velocity}, the RMS velocity of the floaters is compared with $U_o$, showing a linear relation as expected. $U_o$ is an overestimate because is it computed with the maximum magnetic field. Note that the intercept of the linear fit is not zero because the scaling in $U_o$ does not hold at low velocity, where the viscous effects dominate over nonlinearities. A small departure of linearity is also observed at high current. It may be the consequence of important vertical motion occurring at higher driving, both at the surface and in the bulk of the flow. Our horizontal measurement cannot take this motion into account, resulting in a mismatch between the energy injected by the electromagnetic forcing and the one quantified in the measurement. 2D turbulence exhibits an inverse cascade concentrating the energy at the larger scale allowed by the setup. It does not seem to be the case in our flow, which is neither fully turbulent nor fully 2D. In contrast, the averaged vertical vorticity (estimated from floaters motion [22]) exhibits structures of the size of the magnet stripes, as shown in the inset of figure 3. Therefore the forcing length scale $l$, seems indeed the most relevant to describe the turbulent flow. The mean flow contains only a fraction (30 \%) of the total kinetic energy of the flow. One can see the EMDF as a strongly fluctuating flow where vertical vortices of size $l$ compete between each other and interact in an unpredictable way. These vertical vortices induce significant depletion of the surface. At the highest Froude number they can even radiate surface waves. The fluctuations of the surface have a standard deviation around 0.5 mm at $I$ = 200 A and are not Gaussian. Therefore it is very important to perform a coherent phase average to remove all these incoherent fluctuations induced by the EMDF, in order to properly focus on the waves. 

\subsection{General features of the propagating wave}

\begin{figure}
\centering
\vspace{-4cm}
\includegraphics[width= 0.9\columnwidth,angle=0]{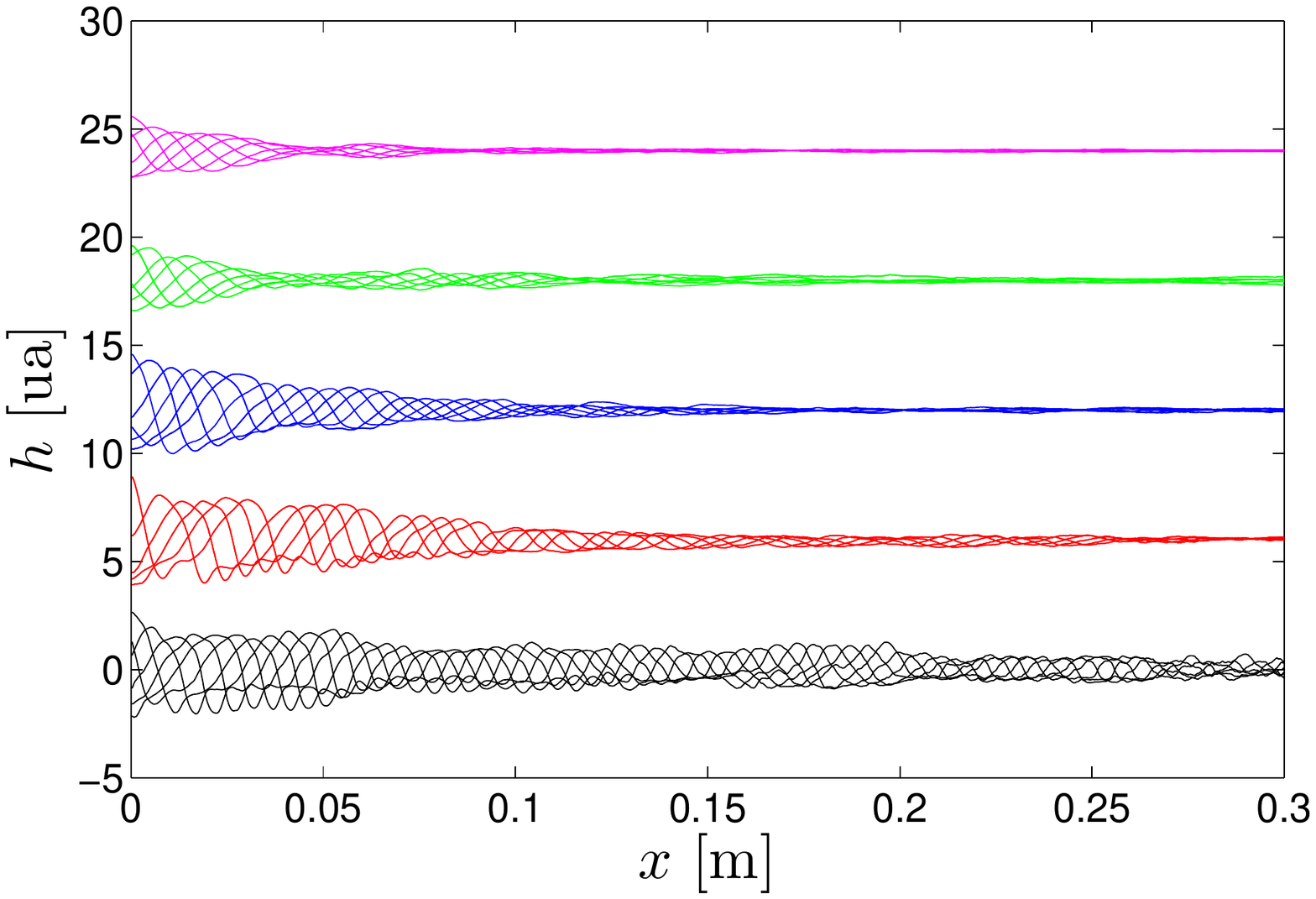}
\vspace{-4cm}
\caption{Spatial profiles of the surface elevation after a coherent phase average at each forcing period. The forcing frequency is 8 Hz and the forcing amplitude $h_o=1.13$ mm without turbulence. The currents  driving the turbulent flows are $I=0$A (black), $I=40$A (red), $I=80$A (blue), $I=120$A (green), $I=160$A (magenta). For each $I$ a vertical displacement has been introduced for clarity.}
\label{TrTps8Hz1}
\end{figure}

\begin{figure}
\centering
\vspace{-4cm}
\includegraphics[width= 0.9\columnwidth,angle=0]{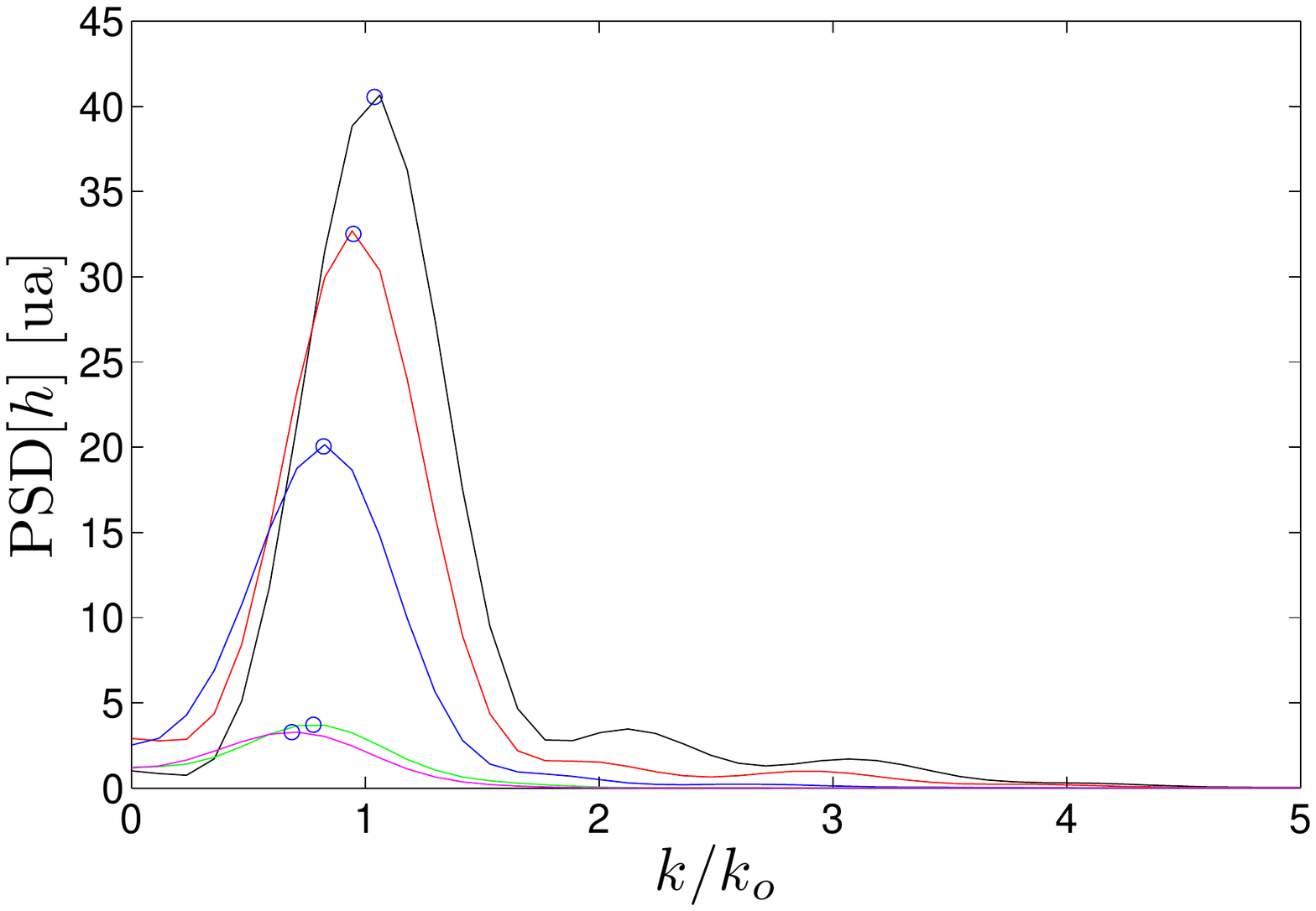}
\vspace{-4cm}
\caption{Power  Spectrum Density (PSD) of the elevation profiles corresponding to those of figure \ref{TrTps8Hz1}. The circles show the peak maximum determined with a polynomial fit.}
\label{PSD8Hz1}
\end{figure}

Figure \ref{TrTps8Hz1} gives an example of the surface deformation induced by the waves after performing a coherent phase averaging. The $n$ averaged curves sampled during a period are plotted for a wavemaker exciting waves at 8 Hz. We can always distinguish an oscillation of the wave with a clear wavelength near the wavemaker chosen as coordinate origin. As the wave goes forward, a damping is visible. The wave propagates on a shorter distance when the supplied current, thus the turbulence strength, is increased. Therefore a damping is induced by the turbulent flow. Figure \ref{PSD8Hz1} shows the Power spectral density (PSD) of the deformation plotted on figure \ref{TrTps8Hz1}. A close look around the main wavenumber peak,  reveals a shift of the wavelength that cannot be seen directly on the averaged profiles. Indeed the abscissa of the PSD maximum (approximated by a polynomial fit) shifts to smaller values when the turbulent intensity is increased. It implies an increase of the wavelength with the increase of the turbulence strength. In the two following parts, (i) we study this wavelength shift  and we try to track some frequency shifts, (ii) we quantify the  turbulent enhancement of the damping. These studies are made for various excitation frequencies and amplitudes of the waves together with various strengths of the turbulence.

\section{Wavelength and frequency shifts}
\label{Shift}
        
\subsection{Shift of the wavelength}
\label{WNDrift}
\begin{figure}
\centering
\vspace{-4cm}
\includegraphics[width= 0.9\columnwidth,angle=0]{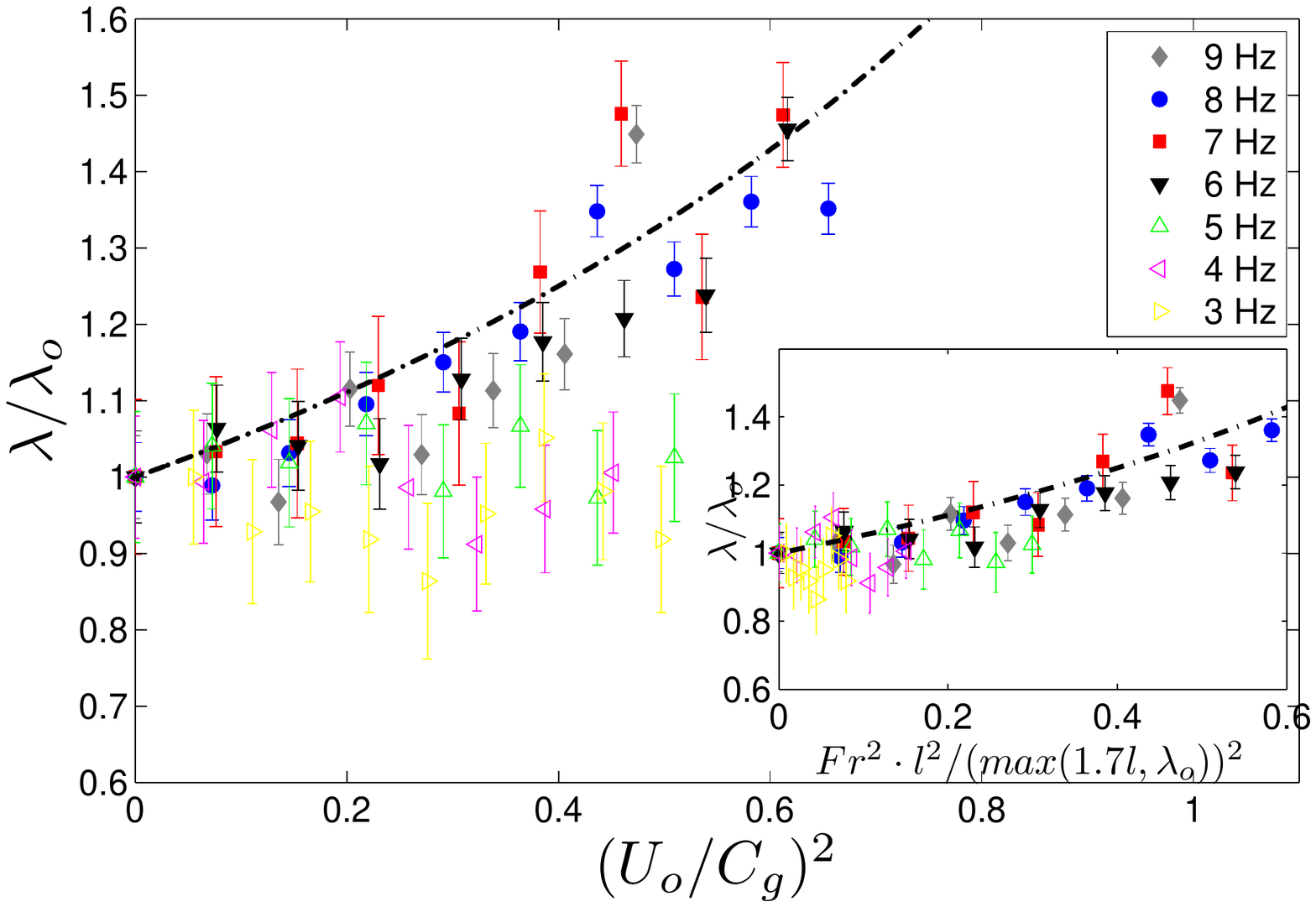}
\vspace{-4cm}
\caption{Wavelength shift induced by the turbulence at various excitation frequencies. The dot-dahed line corresponds to the relation $\frac{\langle  \lambda \rangle }{\lambda_o} =\frac{1}{1-Fr^2}$ (see Equation \ref{Dop}). In the inset,  the x-axis of the lowest excitation frequency (below 5 Hz and therefore $\lambda_o\geq1.7 l$) is weighted by $l/\lambda_o$.}
\label{kvsI}
\end{figure}

In order to study the wavelength shift at all the EMDF and all the excitation frequencies of the wave, we compute the wavelength of the first oscillations of the profile (actually twice the average of the first four half periods). Indeed,  at strong turbulent damping, the oscillating part of the profile is quickly damped. At low excitation frequency only few periods remain. Therefore the study of the  PDS peaks does not yield an accurate determination of the wavelength in this case. We check that both signal processing  coincide to the same wavenumber at large excitation frequency. The main panel of figure \ref{kvsI} shows this effective wavelength as a function of $Fr$ measured at various excitation frequencies of the wave. On the figure \ref{kvsI}, the increase of the wavelength occurs at the excitation frequencies larger than $5$ Hz. Below $5$ Hz the wavelength is nearly constant in our range of Froude  number. We have not noticed significant modification by changing the amplitude of the waves, $h_o$, in the range of forcing tested in our experiment. Hence, the shift seems insensitive to the waves nonlinearity quantified by the steepness: $h_o/\lambda$.
   
We would like to explain this increase of the wavelength at high excitation frequency. In order to do so,  one must recall that the wave  is advected by the underlying turbulent velocity field. Hence, for $\lambda_o\ll l$, with $\lambda_o$ the forced wavelength, the wave does not propagate in a  straight direction. However, we measure experimentally  $k_x$, the wavevector projection on the x-axis of this scattered wave. We assume that the  shift of the wavelength traces the fluctuations of the direction of propagation. Indeed, let us call  $\theta $ the  small  angle of the wavevector with the x-axis. At first order, this deviation induces  a component of the wavevector: $k_y/k_o=sin(\theta)\sim\theta$ and $ k_x/k_o=cos(\theta)\sim 1-\theta^2/2$, where $k_o$ is the wavevector modulus and  $k_x$,  $k_y$ are its projections on the x-axis and the y-axis respectively. The rotation of the wavevector is due to the flow gradients. The wave packet crosses a  flow stucture of size $l_i$ during a time $l_i/C_g$, where the index $i$ labels each structure of the flow. During this time, the wave crest  is deviated about $\Delta y_i=\Delta U_i \cdot l_i/C_g$ where $\Delta U_i$ is the velocity increment induced by the structure $i$. For $\lambda_o \ll l$ the deflection of the wave is given by $\theta_i=\Delta y_i/l_i$.  In average there is no mean flow, then $\langle \Delta k_y\rangle =0$ but $\langle U_i^2\rangle\propto U_o^2$ and 
\begin{equation} 
\frac{\langle  k_x\rangle }{k_o} =1-\frac{Fr_g^2}{2}
\label{Dop}
\end{equation}
 The dashed line shows this scaling on figure \ref{kvsI}. It agrees with the higher excitation frequencies. In this limit, the wavelength shift seems to trace back the fluctuations of the deflection angle of the wave $\langle \theta^2 \rangle$, induced by the flow.

 In the opposite limit $\lambda_o > l$, figure \ref{kvsI} shows that the wavelength is constant. Therefore,  the underlying flow does not deviate the wave anymore. A sharp transition between these two regimes appears on  figure \ref{kvsI} for $\lambda_o/l$ between 1.7 and 1.4.
 Note that in the limit $\lambda_o\gg l$, one may prefer to express the deflection as  $\theta_i=\Delta y_i/\lambda_o= (\Delta U_i\cdot l/C_g)\cdot(\nu_f /C_w)$ where $\nu_f$ is the forcing frequency. Therefore one expects, $\langle  k_x\rangle /k_o =1-(Fr_g\cdot l\cdot \nu_f/C_w)^2/2$ for $\lambda_o\gg l$. If we assume   $\theta_i=\Delta y_i/\lambda_o$ for $\lambda_o> 1.7   l$ and $\theta_i=\Delta y_i/l$ for $\lambda_o< 1.7   l$,  then the  inset of figure \ref{kvsI} shows that all curves collapse together. All the points at large wavelength are sent  near the origin because the control parameter of the deflection is   $Fr_g\cdot l/\lambda_o$ in this limit. This parameter remains small (less than 1/3) in the considered cases.  

\subsection{Frequency shift}
\label{FreqDrift}
This shift of the wave number naturally raises the question about how the frequency changes for a wave propagating through this randomly moving medium.
At small Froude number and in the limit of small wavelength compared to the distance between scatterers, we expect a frequency shift smaller than the one for the wavenumber. Indeed, in this limit, the relation (17) of \cite{Censor75} implies  a Doppler shift of the frequency satisfying $\Delta \omega/\omega_o\sim( C_g/C_w) \cdot (\Delta k/k_o) Fr $ where $\Delta \omega$ and $\Delta k$ are the frequency and wavenumber shifts at a given point and at a given time. Hence the normalized frequency shift is reduced by a factor of the order of the Froude number, compared to the wavenumber shift. For a random velocity field with zero mean, one might have to push the expansion to higher order. Moreover, as we excite the waves at a constant rate, one may expect that the frequency should be fixed close to the paddle.

We  check if there is a frequency shift when the wave travels on the turbulent flow. To do so, we place an inductive sensor at 36 cm from the paddle and 12 cm apart from the propagation axis. Note that such a sensor has a good resolution on the vertical direction but not in the horizontal one. It filters out the small wavelengths. In addition, the fluctuations of the surface induced by the turbulent motion cannot be eliminated by coherent averaging. Consequently, the measurements at low frequencies might be perturbed by the turbulent background and the measurements at large frequencies might be biased by the sensor cutoff.

\begin{figure}
\centering
\vspace{-4cm}
\includegraphics[width= 0.9\columnwidth,angle=0]{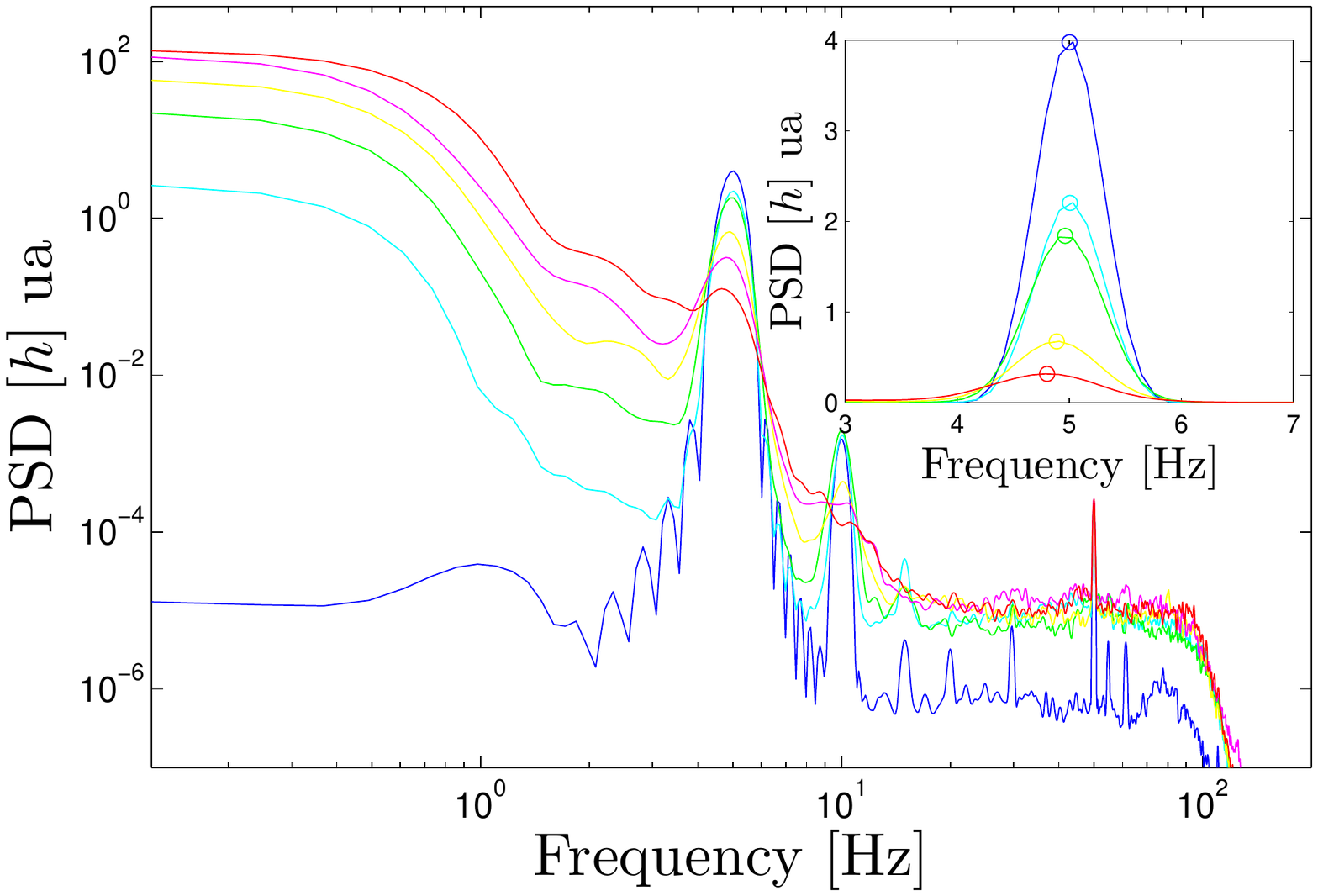}
\vspace{-4cm}
\caption{ The PSD of the local fluctuations in time of the surface elevation for a forcing frequency of 5 Hz. The forcing intensities are $I=0$ A (blue), $I=20$ A (cyan), $I=40$ A (green), $I=60$ A (yellow), $I=80$ A (magenta), $I=100$  A (red). The inset zooms the PSD of the elevation  pass-band filtered around the forcing frequency.}
\label{PDSH_t}
\end{figure}

\begin{figure}
\centering
\vspace{-4cm}
\includegraphics[width= 0.9\columnwidth,angle=0]{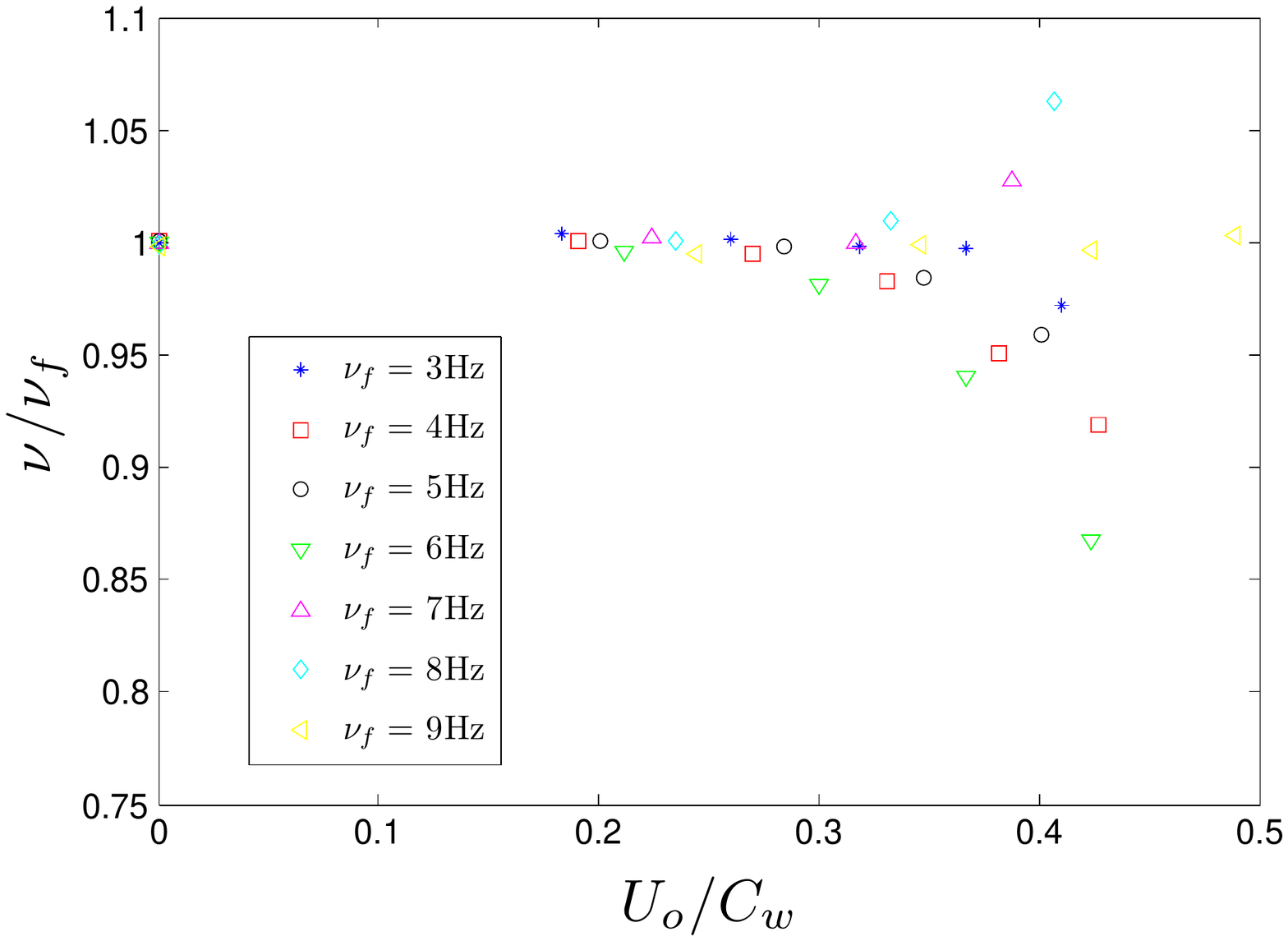}
\vspace{-4cm}
\caption{ Frequency shift of a wave propagating on a turbulent media as a function of the Froude number (based on the phase velocity of the wave). Each symbol corresponds to the different forcing frequencies shown in the legend.}
\label{FreqShift}
\end{figure}

The PSD of the local fluctuations in time of the surface elevation are shown in figure \ref{PDSH_t} for a forcing frequency of 5~Hz. Despite the increase of the Power Spectrum at low frequency, a peak can be seen until $I=80$~A. The bump at low frequency, which increases with the driving, is due to the slow motion of the surface depletions induced by the motion of competing vortices. A closer look of the remaining peak around the excitation frequency reveals a shift. It is underlined in the PSD of the pass-band filtered signal, presented in the inset of figure \ref{PDSH_t}. This shift is shown on figure \ref{FreqShift} for various Froude numbers and various excitation frequencies. Unlike the wavelength shift, the sign of the frequency shift seems to depend on the excitation frequency $\nu_f$ . Two regimes can be identified around $\nu_f$ = 6 Hz. Below this value, the measured frequency decreases with the Froude number. Above $\nu_f=$6 Hz, it seems to be nearly constant or even to increase. The shift is never larger than 15 \% and is weaker than the wavenumber shift, as expected.
We should underline that this shift must depend on the distance travelled by the wave through the turbulent media and on the local properties of the flow. Hence, the shift could evolve with the sensor position. This makes difficult a quantitative comparison between the locally-measured frequency shift and the wavenumber shift evaluated over all the container length. A systematic study of the wavenumber and frequency in space need a much larger device, in order to define unambiguously a local wavelength and a local frequency. This interesting extension of our study is, however, beyond the scope of the present work.

\section{Turbulent Damping}
\label{Damp}

\subsection{Theoretical predictions}
\label{ThPred}
We now focus on the spatial damping of the coherent time-averaged profiles seen in figure \ref{TrTps8Hz1}. We recall that the coherent time averged profiles were obtained by locking the acquisition of the surface profiles  with the excitation frequency. The enhancement of the damping is attributed here to the increase of turbulent fluctuations.  The problem has been considered theoretically from different approaches. Indeed the damping of a progressive wave can be due to the scattering of the wave in others directions by velocity gradients, or be induced by energy exchange with the underlying flow. In this section we first review these predictions and recall their range of validity. The assumptions of the theoretical models never fit exactly our the experimental conditions. Nevertheless, we can check if some theoretical predictions agree with our results, in order to extend these ideas to our experiment conditions. To do so, we emphasize specially the scaling laws between the turbulent damping and the Froude number.

\begin{itemize}

\item Elastic scattering

The damping of sound waves induced by the elastic scattering on turbulent structures has been considered in \cite{Lighthill53}. For sound waves with a wavelength less than the Taylor microscale, $\lambda_T$, a damping $\gamma_{turb}\propto (\langle v^2 \rangle/c^2)/\lambda_T$ is suggested, with $c$ the sound speed. Phillips used the same kind of elastic scattering arguments to determine the damping of surface gravity waves by turbulence \cite{PhillipsJFM59}.  In the oceanographical context he considered, the wavelength of the gravity waves is smaller than the forcing length scale of the turbulence. Moreover, he restricted his analyses to the {\it single scattering limit} and the Born approximation. This limit is valid only when the scatterers are sparse. In this case, the complete wave field can be approximated by the incident wave in the scattering process. It requires a wave energy much larger than the turbulent energy. In terms of Froude number it imposes $Fr_w\ll h_o/\lambda$. This approach has been extended later to drift flow turbulence by Fabrikant and Reavsky \cite{FabrikantRaevskyJFM93}. In our experiment, we keep the wave steepness $h_o/\lambda<0.1$ in order to prevent wave breaking. The second and less demanding requirement is that the duration vorticity fluctuations are small compared to the time spend by the wave inside the turbulent area. This imposes $l/L\ll Fr$. In our experimental cell one has always $Fr > h_o/\lambda$. Hence, we do not seem in the single scattering limit. Therefore the  predictions of Phillips, giving $\gamma_{turb}/k=(k\cdot l)^{-2/3}Fr^2$, do not  seem  the most pertinent to interpret our experimental results.

However  in section \ref{WNDrift},  we have interpreted  the wavelength shift as an elastic deflection of the wave by the flow, in the limit of small wavelength. We have established that the variance of the deflection angle is  $\langle \theta^2\rangle=Fr^2$. We can infer a {\it total scattering cross section} of the wave (actually a line in our case of surface waves)  from this deflection: $\sigma*=a(1-\langle cos(\theta)\rangle)\sim a\cdot Fr^2$, where $a\leq l$ would be associated to the mean size of the vortex core. Therefore, we know formally: the density of vortices $n\sim1/ l^2$, the mean free path of the wave $l^*= l^2/(a\cdot Fr^2)\leq l/Fr^2$ and then the number of scattering events during the crossing of the cell, $N^*=L/l^*\leq Fr^2\cdot L/l\approx 4$. It proves that we are not in the multi-scattering regime in which the waves follow a diffusive random walk in the disordered media. Hence, if the damping is mainly due to elastic scattering, we can approximate the damping coefficient by $ 1/l^*$ hence $\gamma_{turb}l \propto Fr^2\cdot a/l$. Concerning the scaling in $Fr$, this crude estimation agrees with the Phillips, Fabrikant and Reavsky predictions. However, these  authors have shown that the scattering of surface gravity waves by a vortex is a complex phenomena and the scattering cross section is far to be isotropic \cite{FabrikantRaevskyJFM93}.
  
\item Vortex stretching

More recently, Teixeira and Belcher studied the opposite limit where the wavelength is larger than the integral scale of the turbulence \cite{TeixeiraJFM02}. They used the Rapid Distortion Theory requiring $\lambda\gg l$ and $Fr\ll h_o/\lambda$.
In this framework, they proposed a mechanism of energy transfer from the wave to the turbulence, via a stretching of vortices induced by the Stokes drift of the surface wave. At the end, they obtained, $\gamma_{turb}/k=\langle\delta U_o^2\rangle/C_g^2$.

One must note that despite the different mechanisms and limits considered in \cite{PhillipsJFM59} and \cite{TeixeiraJFM02}, both approaches end with the same kind of scaling law. Therefore the scaling in Froude number is not enough to infer the underlying  physical mechanisms. Indeed, Phillips found a damping:
\begin{equation}
\gamma_{turb} \propto \frac{k^2}{C_g^2}\int_{-\pi}^{\pi} \Phi({\mathbf \kappa}) d\theta,
\label{Phillips59}
\end{equation}
where ${\bf \kappa}$ is the scattered wavevector, $\theta$ the scattered angle and $ \Phi({\bf \kappa})$ is the Power Spectrum of the vertical vorticity. Dimensionally, it implies a damping $\gamma_{turb}=k \langle\delta v_k^2\rangle/C_g^2$, where $\delta v_k=U_o(l\cdot k)^{-1/3}$ is the typical turbulent velocity increment on a scale $k$. Indeed in the single scattering limit, the velocity increments on a scale of order of the wavelength are the most involved in the damping process. Teixeira and Belcher exhibited a damping process of the wave energy, $E_w$,  involving the turbulent Reynolds stress proportional to $U_o^2$ times the velocity gradient of the Stokes drift proportional to $h_o^2 k^2\omega$. One gets : $\frac{1}{E_w}\frac{dE_w}{dt}\sim\frac{U_o^2\cdot(h_o^2 k^2\omega)}{(h_o\omega)^2}$ and $\gamma_{turb}\propto  \frac{1}{C_g\cdot E_w}\frac{dE_w}{dt}\sim k (U_o/C_g)^2$. Therefore, both approaches scale with the square of the Froude number. They only differ in the wavenumber dependency, because they involve a different amount of  turbulent kinetic energy. Unfortunately, as in most of the laboratory size experiments, we are in the crossover region where the wavelengths are of order of the forcing scale. Moreover the  requirement that $Fr\ll h_o/\lambda$, shared  by the two models,  is not achieved in our setup. 

\item Turbulent advection

Boyev considered a case where the energy of the turbulent flow is much larger than the energy carried by the wave. It is a relevant asssumption for our experimant. He also assumed that the turbulent time scale is much smaller than the wave period to facilitate an efficient mixing of the energy \cite{Boyev71}. The former assumption imposing $h_o/\lambda\ll Fr$ is nearly true in our experiment. The latter assumption implies $l/\lambda\ll Fr$ which is clearly not satisfied in our device. In the deep water model proposed by  Boyev, the energy in the wave is limited into a small layer (of order $1/k$). This energy of the wave is advected into the bulk of flow at the turbulent mixing rate, $\Omega_m$.  Following the estimation of the mixing rate across a layer of size $1/k$ proposed in \cite{OlmezJFM92}: $\Omega_m=U_o k^{2/3}/l^{1/3}$, one gets

\begin{equation}
\gamma_{turb}\propto \frac{ U}{C_g(k)} k\cdot (l\cdot k) ^{-1/3}.
\label{gammaturv2}
\end{equation} 
This mechanism, transferring the energy of wave into the deep turbulence, is not efficient in our layer of Gallinstan where the depth is small and where the vorticity is mainly vertical. However one can assume that the same mechanism is at work for our progressive waves. Indeed, a horizontal mixing rate can scatter the waves and extract the energy from the direction of propagation.

\end{itemize}

\subsection{Experimental results}

There are many ways to characterize the damping from the profiles shown in figure (\ref{TrTps8Hz1}), but all must be equivalent. We choose to compute the standard deviation of the profile. Assuming an exponential turbulent decay with an damping coefficient $\gamma_{turb}$, one has a {\it Damping enhancement}: $\gamma_{turb}/\gamma_o\propto\sqrt{\langle \Delta h_o^2 \rangle}/\sqrt{\langle \Delta h_{turb} ^2\rangle}$, where $\Delta h$ is the surface deformation induced by the wave and where the subscripts $o$ and $turb$ correspond respectively to the cases without and with turbulence. Assuming that all viscous effects (including purely magnetic damping, as discussed below), are included in $\gamma_o$, one expects $\gamma_{turb}/\gamma_o\propto F(U/C_g, h_o/\lambda_o,k_o\l,C_w/C_g)$. In our range of parameters, the ratio of the phase velocity over the group velocity does not evolve a lot (see Tab 1), hence it can be excluded from the scaling law. At a given frequency, we note that a better collapse of the curve $\gamma_{turb}/\gamma_o$ vs $U/C_g$ is obtained if the damping enhancement is multiplied by the wave steepness $h_o/\lambda_o$. It also improves the data collapse at high frequency (above 6 Hz). The benefit is less obvious at lower frequencies. Note that the higher frequencies do not correspond necessarily to the higher steepness since we vary also the forcing amplitude. Moreover we can also express the damping ratio $\gamma_{turb}/\gamma_o$ as a function of the combination: $Fr_g\cdot l/\lambda_o$. Indeed this parameter can be interpreted as the ratio of two charateristic timescales: (i) $\tau_w=l/C_g$, the time spent by the wave to cross a turbulent structure,  (ii) $\tau_m=\lambda_o/U_o$, the horizontal turbulent mixing rate over a size $\lambda_o$.  It can be understand as the time used by the turbulent flow to move the fluid about a wavelength.

\begin{figure}
\centering
\vspace{-4cm}
\includegraphics[width= 0.9\columnwidth,angle=0]{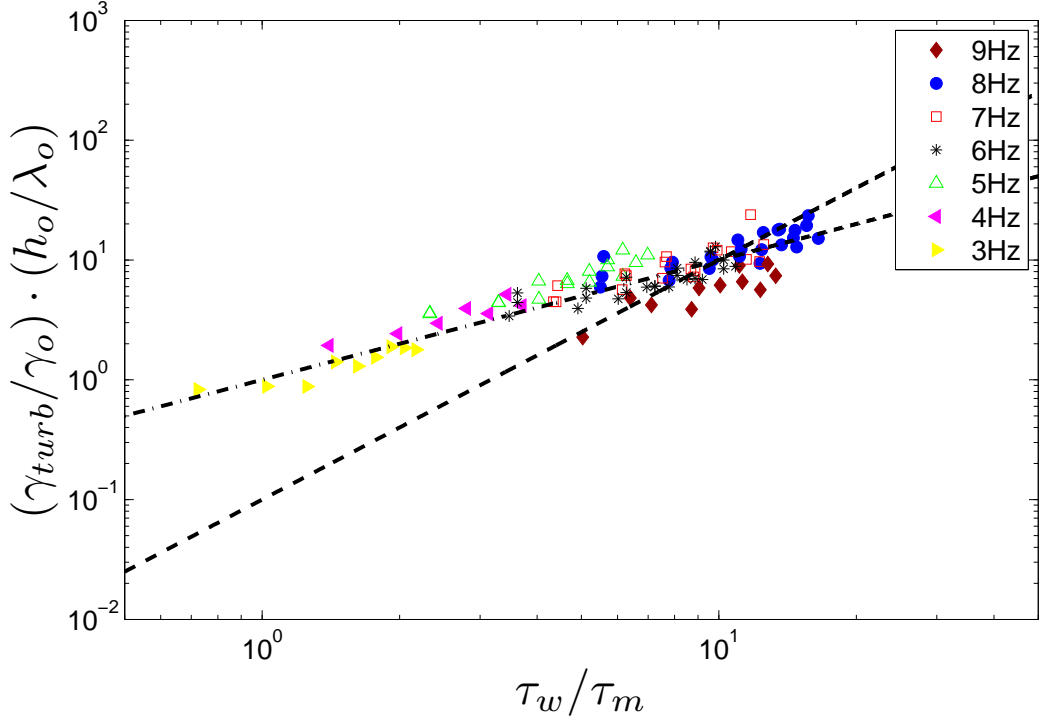}
\vspace{-4cm}
\caption{Damping  induced by the turbulence as a function of the ratio between the time spent by the wave to cross a turbulent structure, $\tau_w$, and the turbulent mixing time over a length $\lambda_o$ , $\tau_m$ with $\tau_w/\tau_m=Fr_g\cdot l/\lambda_o$, for various excitation frequencies and amplitudes. Dashed line corresponds to $\gamma_{turb}\propto (U/C_g)^2$ , and the dot-dashed line  corresponds to $\gamma_{turb}\propto (U/C_g)$.}
\label{DampPlot}
\end{figure}

Figure \ref{DampPlot} shows the variations of the turbulent damping enhancement multiplied by the steepness as a function of the time ratio $\tau_w/\tau_m=Fr_g\cdot l/\lambda_o$ in logarithmic representation. We choose these parameters to underline the role of the two timescales $\tau_m$ and $\tau_w$. It comes out from our data that the best fit is linear (dot-dashed line) in agreement with the scaling proposed in \cite{Boyev71}. The dashed-line represent a scaling with  power 2, which  gives a less good fit. This linear behavior in terms of mixing rate promotes a damping mechanism by turbulent advection \cite{Boyev71,OlmezJFM92}.  However, in our experimental configuration, it is most probable that the energy of the waves depart from its propagation direction because of a {\it horizontal advection} by the vertical vorticity. The linear elastic scattering that can explain the wavelength shift, would involve the square of the Froude number. Hence it is not the main mechanism implied in the damping. The action of the wave amplitude remains puzzling. It may trace the role of the nonlinearity (proportional to the steepness) as suggested in  \cite{GeenNatPhysSci72} or it may underline the Stoke drift of the wave (having a velocity proportional to $C_w(h_o/\lambda)^2$). 

Finally,  we have not yet consider pure MHD actions regarding the damping. Howerver, without applied currents, it has been shown that the wave motion above a magnetic field  induce Foucault currents that dissipate energy by Joule effect and hence damp the waves \cite{Etay}. This effect is taken into account in $\gamma_o$. When an applied current is added, the full problem implies: surface deformation by the wave and the flow and induction of inhomogeneous magnetic field. It is difficult to do predictions in this very complex configuration.  One may assume that the wave might modulate the EMDF by modulating the Hartmann layer for instance. These might induce resonances when the wavelenght matches with the mesh of the magnet array. Hence an MHD effect cannot be excluded concerning the  role of the amplitude in the damping enhancement. This would deserve further theoretical  studies.

\section{Discussions and Conclusions}

Our work is devoted to the study of a propagating wave above an almost 2D flow. Our experimental setup has a large ratio of  turbulent energy over the wave energy. We observe a significant shift of the component of the wavenumber along the propagation direction, as long as  the excited wavelength is smaller than the forcing length scale of the turbulent flow. In this case, the value of the projection of the wavevector along the main propagation axis decreases when the Froude number is increased. To our knowledge, this has not been reported before. This shift is interpreted as the signature of the fluctuations on the direction of wave propagation. Our  argument assumes that the modulus of the wavevector is conserved during the wave propagation on the turbulent flow. However, a shift of the wave frequency is also observed far from the forcing point. These results raise the question about how the dispersion relation of surface waves is affected  by the  presence of an underlying complex flow. Indeed the dispersion relation (\ref{reladisp}) is deduced only for a potential flow. Otherwise the separation of the flow into a surface wave and a background flow is questionable. Some procedures are suggested in \cite{Kitaigorodskii2}, but no modification of the dispersion relation is proposed by these authors. Dispersion relations have been computed recently in some specific cases with vertical vorticity and only along a single horizontal direction (1D-waves) \cite{Constantin,Martin}.   To go further in this direction, a complete study of the surface deformation in the frequency-wavenumber space would be useful. The wavemaker could be driven by a random excitation in order to generate several frequencies and wavenumbers. However, we do not have yet a spatial and temporal resolution accurate enough to determine the dispersion relation in the presence of randomly moving fluid. Another difficulty would be to remove accurately the deformation induced by the flow. Finally, at high Froude number, the flow should generate waves by itself \cite{vandeWater}.

A second point of our study concerns the damping of the wave. We show a turbulent damping increasing with the Froude number. The best fit is linear with the Froude number, as proposed in \cite{Boyev71, OlmezJFM92}. The observed enhancement of the damping is not compatible with the quadratic scaling of Phillips, Fabrikant and Raevsky and of Teixeira and Belcher \cite{PhillipsJFM59,FabrikantRaevskyJFM93,TeixeiraJFM02}. The linear scaling can be interpreted  as a consequence of a horizontal mixing of the energy of the progressive waves. Two scenarios remain possible (i) the energy is deviated but remains as surface waves, or (ii) it is mainly absorbed in the bulk of the turbulent flow. The former case conserves the  energy of the surface wave although the energy can be transfered to other waves by nonlinearities. In the second case, the energy is  directly  exchanged with the flow and the scattering of the waves is strongly inelastic. To discriminate between these two mechanisms, a 2D measurement of the entire surface deformation would be helpful. However, it will be hardly done with a liquid metal. Moreover, the electromagnetic forcing may induce complex effects which remain unclear at present. Nevertheless, one advantage of our device is that we can generate various flows. For intance we force here mainly vertical vortices, but we can also generate horizontal vortices aligned with the wavetrain, by rotating the magnets by 90$^o$. We expect a very different action of the flow in this case. This modification should help to distinguish between the damping due to elastic scattering from the one induced by advection of the wave energy. Moreover a measure of the mean power injected by the wavemaker as function of the tubulence strength could give also a new insight to this question.  Such studies are planned to understand the mechanisms of energy exchange on fully 3D turbulent flows.

\begin{acknowledgments}
We would like to thank V. Padilla for building the setup, C. Wiertel--Gasquet for helping at the automation of the experiment, B. Gallet, N. Mujica, A. Tanguy, M. Berhanu  and M. Bonetti for helpful discussions  and  useful comments. This work is supported by the ANR Turbulon. PG also received support from the triangle de la physique and CONICYT/FONDECYT postdoctorado N$^o$ 3140550.
\end{acknowledgments}


\begin{thebibliography}{1}
\bibitem{Sheng}
P. Sheng, {\it Introduction to Wave Scattering Localization and Mesoscopic Phenomena} (Academic Press, San Diego, CA, USA, 1995)

\bibitem{Ishimara}
A. Ishimara, {\it Wave propagation and scattering in Random Media}, (IEEE Press, New York, NY, USA, 1997)

 \bibitem{Phillips77}
O. M. Phillips, {\it Dynamics of the upper Ocean} (Cambridge University Press, 2d Ed., 1977)

\bibitem{Kitaigorodskii1}
S. A. Kitaigorodskii, M. A. Donelan, J. L. Lumley, and  E. A. Terray, ``Wave-Turbulence Interactions in the Upper Ocean. II: Statistical Characteristics of Wave and Turbulent Components of the Random Velocity Field in the Marine Surface Layer," J. Phys. Oceanogr. {\bf 13}, 1988--1999 (1983).

\bibitem{Ardhuin2006}
F. Ardhuin and A. D. Jenkins, ``On the Interaction of Surface Waves and Upper Ocean Turbulence," J. Phys. Oceanogr. {\bf 36}, 551--557 (2006).

\bibitem {Veron2009}
F. Veron, W. K. Melville, and L. Lenain, ``Measurements of Ocean Surface Turbulence and Wave--Turbulence Interactions," J. Phys. Oceanogr. {\bf 39}, 2310--2323 (2009).

\bibitem{VivancoPRE04}
F. Vivanco and F. Melo, ``Experimental study of surface waves scattering by a single vortex and a vortex dipole," Phys. Rev. E {\bf 69}, 026307 (2004).

\bibitem{CostePRE99}
Ch. Coste, F. Lund, and M. Umeki, ``Scattering of dislocated wave fronts by vertical vorticity and the Aharonov-Bohm effect. I. Shallow water," Phys. Rev. E {\bf 60}, 4908--4916 (1999); Ch. Coste and F. Lund, ``Scattering of dislocated wave fronts by vertical vorticity and the Aharonov-Bohm effect. II. Dispersive waves," Phys. Rev. E {\bf 60}, 4917--4925 (1999).

\bibitem{Lighthill53}
M.J. Lighthill, ``On the energy scattered from the interaction of turbulence with sound or shock waves," Proc. Camb. Phil. Soc. {\bf 49}, 531--551 (1953). 

\bibitem{PhillipsJFM59}
O. M. Phillips, ``The scattering of gravity waves by turbulence," J. Fluid Mech. {\bf 5}, 177--192 (1959).

\bibitem{FabrikantRaevskyJFM93}
A. L. Fabrikant and M. A. Raevsky, ``Influence of drift flow turbulence on surface gravity wave propagation," J. Fluid Mech. {\bf 262}, 141--156 (1994).

\bibitem{TeixeiraJFM02}
 M. A. C. Teixeira and  S. E. Belcher, ``On the distortion of turbulence by a progressive surface wave," J. Fluid Mech. {\bf 458}, 229--267 (2002).

\bibitem{Kantha2006}
L. Kantha, ``A note on the decay rate of swell," Ocean Model. {\bf 11}, 167--173 (2006).

\bibitem{Boyev71}
A. G. Boyev, ``The damping of surface waves by intense turbulence," Izv., Atmos. Ocean. Phys. {\bf 7}, 31--36 (1971).

\bibitem{GuoShen2010}
X. Guo and L. Shen, ``Interaction of a deformable free surface with statistically steady homogeneous turbulence," J. Fluid Mech. {\bf 658}, 33--62 (2010). 


\bibitem{GuoShen2014}
X. Guo and L. Shen, ``Numerical study of the effect of surface wave on turbulence underneath. Part 2. Eulerian and Lagrangian properties of turbulence kinetic energy," J. Fluid Mech. {\bf 744}, 250--272 (2014).

\bibitem{Claudio}
C. Falc{\'o}n and S. Fauve, ``Wave-vortex interaction", Phys. Rev. E {\bf 80}, 056213 (2009).

\bibitem{GeenNatPhysSci72}
T. Green, H. Medwin, and J. E. Paquin, ``Measurements of surface wave decay due to underwater turbulence," Nature Phys. Sci. {\bf 237}, 115--117 (1972).

\bibitem{OlmezJFM92}
H. S. Olmez and J. H. Milgram, ``An experimental study of attenuation of short water waves by turbulence," J. Fluid Mech. {\bf 239}, 133--156 (1992).

\bibitem{Ermakov2014}
S. A. Ermakov, I. A. Kapustin, and O. V. Shomina, ``Laboratory Investigation of Damping of Gravity-Capillary Waves on the Surface of Turbulized Liquid," Izv., Atmos. Ocean. Phys.  {\bf 50}, 204--212 (2014).

\bibitem{Gallinstan}
From the safety datasheet acc, Guideline 93/112/EC of Germatherm Medical AG,  the Gallinstan is made of 68.5 \% of Gallium, 21.5 \% of indium, 10 \% of Tin. Its density is
$\rho=6.440\times 10^3$~kg/m$^3$, its dynamic viscosity is
$\nu_G=3.73\times 10^{-7}$~m$^2$/s, its electrical conductivity
$\sigma=3.46 \times 10^6$~S/m.

\bibitem{GutierrezTh}
Gutierrez P. PhD Thesis , ``Effects on the free surface of a turbulent flow'',  Ecole Polytechnique, 2013 

\bibitem{Young}
Y. K. Tsang and W. R. Young, ``Forced-dissipative two-dimensional turbulence: A scaling regime controlled by drag," Phys. Rev. E {\bf 79}, 045308(R) (2009). 

\bibitem{Sommeria_1986}
J. Sommeria, ``Experimental study of the two-dimensional inverse energy cascade in a square box," J. Fluid Mech. \textbf{170}, 139--168 (1986).

\bibitem{Censor75}
D. Censor, ``The group doppler effect," J. Franklin Inst. {\bf 299}, 333--338 (1975). 

\bibitem{Kitaigorodskii2}
S. A. Kitaigorodskii and J. L. Lumley 
``Wave-Turbulence Interactions in the Upper Ocean. I: The Energy Balance of the Interacting Fields of Surface Wind Waves and Wind-Induced Three-Dimensional Turbulence," J. Phys. Oceanogr. {\bf 13}, 1977--1987 (1983).

\bibitem{Etay}
B. Seernivasan, P.A. Davidson, J. Etay, ``On the control of surface waves by a vertical magnetic field'',  Phys  Fluids {\bf 17}, 117101 (2005)

\bibitem{Constantin}
A. Constantin, ``Dispersion relations for periodic traveling water waves in flows with discontinuous vorticity," Comm. Pure Appl. Anal. {\bf 11}, 1397--1406 (2012).

\bibitem{Martin}
C. I. Martin, ``Dispersion relations for periodic water waves with surface tension and discontinuous vorticity," Disc. Cont. Dyn. Syst. {\bf 34}, 3109--3123 (2014).

\bibitem{vandeWater}
R. Savelsberg and W. {van de Water}, ``Turbulence of a Free Surface," Phys. Rev. Lett. {\bf  100}, 034501 (2008).
\end{thebibliography}
\end{document}